\documentclass[10pt,journal,onecolumn]{IEEEtran}
\IEEEoverridecommandlockouts
\usepackage{cite}
\usepackage{amsmath,amssymb,amsfonts}
\usepackage{algorithmic}
\usepackage{graphicx}
\usepackage{textcomp}
\usepackage{xcolor}
\usepackage{csquotes}
\usepackage[bookmarks=false]{hyperref}
\usepackage{tabularx,multirow,booktabs,blindtext}
\def\BibTeX{{\rm B\kern-.05em{\sc i\kern-.025em b}\kern-.08em
    T\kern-.1667em\lower.7ex\hbox{E}\kern-.125emX}}
\begin{document}

\title{Launching Stealth Attacks using Cloud\\
}

\author{\IEEEauthorblockN{Moitrayee Chatterjee\IEEEauthorrefmark{1}, Prerit Datta\IEEEauthorrefmark{1}, Faranak Abri\IEEEauthorrefmark{1}, Akbar Siami Namin\IEEEauthorrefmark{1}, and Keith S. Jones\IEEEauthorrefmark{2}}
\IEEEauthorblockA{\IEEEauthorrefmark{1} \\
\textit{Department of Computer Science}, \IEEEauthorrefmark{2} 
\textit{Department of Psychological Sciences}\\
\textit{Texas Tech University}\\
\{moitrayee.chatterjee, prerit.datta, faranak.abri, akbar.namin, keith.s.jones\}@ttu.edu}
}

\maketitle

\begin{abstract}
Cloud computing offers users scalable platforms and low resource cost. At the same time, the off-site location of the resources of this service model makes it more vulnerable to  certain types of adversarial actions. Cloud computing has not only gained major user base, but also, it has the features that attackers can leverage to remain anonymous and stealth. With convenient access to data and technology, cloud has turned into an attack platform among other utilization. This paper reports our study to show that cyber attackers heavily abuse the public cloud platforms to setup their attack environments and launch stealth attacks. The paper first reviews types of attacks launched through cloud environment. It then reports case studies through which the processes of launching cyber attacks using clouds are demonstrated. 
We simulated various attacks using a virtualized  environment, similar to cloud platforms, to identify the possible countermeasures from a defender's perspective, and thus to provide implications for the cloud service providers. 
\end{abstract}


\section{Introduction}
Cloud computing employs virtualization to provide users with computing assets on demand, including data, processor, memory, network bandwidth, security services, operating platforms, software, and hardware clusters. Users can enable this access to computing resources through the Internet and achieve flexibility with respect to the resources and their requirements at an affordable cost.
On the flip side, the lucrative features of cloud computing have received much considerations from cyber attackers. The adversaries are increasingly abusing the affordable resources and the security flaws of cloud computing  to stay  \enquote{\textit{stealth}} and launch attacks. 

The cloud-based attacks are becoming prevalent, especially the ones comprising data ex-filtration and information leakage, owing to insufficient security measures, credentials saved on public source code repositories, and the use of weak passwords, to name a few. The security reports published by the public cloud providers\footnote{Microsoft Security Intelligence Report, Volume 22, January through March, 2017}, and our study presented in this paper, indicates the incessant abuse of cloud platforms for launching cyber attacks. The 2017 Microsoft Security Intelligence Report$^1$, reports \enquote{weaponizing} the cloud through creating or gaining access to VMs and launching attacks. Once the attackers are on the cloud, they can launch brute force attacks, propagate spams or run malicious programs and scan cloud-based systems for detecting any vulnerability to exploit. 

The Google Cloud Platform (GCP) has previously reported of being abused for launching DoS and intrusion attacks\footnote{https://www.gcppodcast.com/post/episode-47-cloud-abuse-with-swati-and-emeka/}. Furthermore, attackers have used GCP for \textit{crypto-jacking} and hosting copyright-protected items. The Cloud Security Alliance \cite{c9} has flagged the \enquote{abuse and nefarious} use of the {\it Infrastructure-as-a-Service} (IaaS) as the highest security concern of cloud platform.  
While, the abuse of the cloud may benefit the attackers to remain stealth and do not impact the service provider directly, our study indicates that cloud providers need to tighten their user authentication process and be more proactive in tracking malicious activities on any cloud account in order to prevent the cloud from being abused as a  launching platform for performing any stealth cyber attacks.

In our earlier work on cloud abuse \cite{c15-0}, we provided a list of recommendations for cloud providers in order to tighten their security controls on cloud. This paper complements our initial work from various aspects. This paper reports the abuse detection of cloud, which is complementary study of the interviews responses of the security professionals and ethical hackers, who participated in the professional hacking conferences such as DEF CON and Black Hat. We interviewed 75 professional hackers and discovered that attackers are increasingly abusing the resources on cloud for setting up their attack environments that is not only  cost effective, but also enables them to remain stealth while executing the steps of cyber kill chains\footnote{https://www.lockheedmartin.com/content/dam/lockheed-martin/rms/documents/cyber/LM-White-Paper-Intel-Driven-Defense.pdf}.
The paper highlights the mitigation strategies to counteract such cloud-based attacks. The key contributions of this paper are:
\begin{itemize}
    \item 
     Presenting a holistic analysis of the cloud abuse from the perspective of attackers, representing the {\it ``mental model''} of attackers while launching attacks. 
    \item Simulating different generic attack steps that are performed by attackers and inspecting various log files to identify areas to deploy detection mechanisms for the attack VMs. 
\end{itemize}
The rest of this paper is organized as follows: we explain the motivation and purpose of this study in Section \ref{sec:motivation}. Section \ref{sec:state} presents the state-of-the-art on how cloud is being abused for launching cyber attacks. 
We simulated 3 different attacks on a virtualized environment and provided details of how to capture those activities in Section \ref{sec:experiment}. The related work are reviewed in Section \ref{sec:relwrk}. Finally, we conclude the paper in Section \ref{sec:conclude} and provide some insights about the future research directions. 

\section{Motivation: Attacker's Mental Model}
\label{sec:motivation}

The research team recruited over 75 
security professionals and ethical hackers as participants at the professional hacking conferences (DEF CON and Black Hat) for the purpose of a larger project with the goal of analyzing attacker's mental model while launching an attack. 
The interviewers, who were part of the research team and also graduate students, presented each of the participants with attack scenarios with a very generic hypothetical setting. The participants were approached randomly and were asked if they had any prior hacking experience and if so, they were asked if they would like to participate in the research study. Table \ref{tab:scenarios} lists the attack scenarios that were presented to the participants.
\begin{table*}[t]
\caption{The attack scenarios presented to the participants.}
\label{tab:scenarios}
\begin{tabular}{p{17cm}}
\hline
Scenario 1. Changing Contents of a Website: \\
\textit{You want to boost your own small business by changing the ranking (1 – 5 stars) recorded by the customers who have been the clients of your business. You want to modify the content and make its ranking and reputation great (e.g., changing 1 star to 5 stars). A Website records the ranking entered by the clients of the business. }\\
\hline
Scenario 2. Data Tampering: \\
\textit{Your close friend who is working for a company is not happy with his salary. He asks you to enter the company’s Website and increase his salary by giving you his user name and password. The company has an online payment system.}\\
\hline
Scenario 3. Denial of Service: \\
\textit{There is a competition between the Dog-lovers and the Cat-lover’s parties for the up-coming election. As a cat-lover, you decide to take the main site of the dog-lover down, even for a small amount of time. }\\
\hline
Scenario 4. Deleting/Stealing Internet Usage and Data: \\
\textit{You heard that your Internet provider company would be selling user data and usage habits to advertisers soon. You are obsessed with your privacy and are anger of having our data sold to the third-party. You decide to penetrate to their system and remove your usage data and Internet habits.} \\
\hline
Scenario 5. Email Account Information: \\
\textit{You suspect that your girlfriend is cheating on you! She uses RocketMail, can you determine if she has been exchanging some love emails with her secret lover? Her email address is loveseeker@rocketmail.com.}  \\
\hline
Scenario 6. Open-Ended:\\
\textit{If the participant wants to share their experience in launching a cyber attack(s) that is not covered by the above scenarios.} \\
\hline
\end{tabular}
\vspace{-0.12in}
\end{table*}

The participants were asked to choose one of the scenarios from the list and then describe their approach on how to launch the underlying attack. 
The research team and interviewers did not collect any demographic or personal information, so that the participants can be unguarded about sharing their knowledge and skills as a professional hacker without the apprehension of being exposed. The only question we asked after presenting the scenario was:
\textit{How would you do the attack described in the scenario?}
As the interview progressed, the research team asked the participants probing and follow-up questions to better understand their perspective and comprehend their mental models. The research team collected the responses on paper and manually transcribed them into use cases during the analysis. Table \ref{tab:sampleUC} presents a sample use case that is transcribed using one of the interview responses. 

\begin{table*}[t]
\caption{A use case for tampering ISP usage and data.}
\label{tab:sampleUC}
\begin{tabular}{p{17cm}}
\hline
{\it Use Case:} Tampering ISP usage and data \\
\hline
{\it Primary Actor:} An Attacker \\
\hline
{\it Precondition:} \\
1. Attacker has successfully created an account on cloud and has the computing instance ready for use. \\
2. Attacker has basic knowledge of the ISP server. \\
3. Attacker has necessary network and domain access. \\
4. Attacker has necessary skills and expertise to perform scan and construct malicious scripts.\\
\hline
{\it Description:} \\
1. Create a VPS on AWS instance. \\
2. Setup multi-hop VPN.\\
3. Encrypt channel. \\
4. Use tor browser.\\
5. Set up tools for scanning or developing malicious payloads.\\
6. Scan open ports and interfaces on ISP server for credentials.\\
7. Construct $SQL_i$ script or log in to database. \\
8. Launch $SQL_i$ attack or change the database content.\\
\hline
{\it Post Condition:} \\
Attacker is able to access the database of ISP and delete or modify the required information.\\
\hline
\end{tabular}
\vspace{-0.12in}
\end{table*}

Next, each of the use cases was analyzed for the purpose of discovering patterns. The objective was to build a general mental model of attackers to elicit their thought process during the attack process, which will eventually help in guiding cyber defense personnel in preparation for similar attacks. 

While looking for common patterns in the transcribed use cases, the research team discovered that the attackers are extensively using the publicly available infrastructures, including the cloud, for hosting their attack artifacts. Based on the interview responses, the use cases helped to analyze and build an exhaustive sequence of actions an attacker performs to establish the backbone for launching an attack. 
By enumerating the use cases, we can ascertain how cyber attackers misuse the cloud and further; we can propose the solutions and mitigation to prevent the abuse of cloud.

\section{Attack Types Launched on Cloud}
\label{sec:state}

The use of the cloud for conducting malicious activities is turning out to be one of the biggest challenges in the cloud platform. According to the 2017 cloud security alliance (CSA) report \cite{c9}, a group of attackers was able to successfully use the Amazon AWS cloud service to launch a Distributed Denial-of-service (DDoS) attack. In another report published by the 2017 Microsoft Security Intelligence report \cite{c2}, about 51\% of attacks, in which cloud on Microsoft's Azure platform was used, corresponds to interactions with an external malicious IP address.  These malicious IP addresses are capable of sending further instructions to compromise the security of the cloud. Furthermore, 23\% of the attacks involved performing brute force attacks against scanning remote desktop protocol (RDP) ports on target systems to gain administrative-level access control to the victim systems \cite{c11}. In addition, over 19\% of the attacks involved using the cloud for spamming.  


According to the definition of cloud computing \cite{cloudnist} provided by NIST, there are three primary cloud service models: (1) Software as a Service (SaaS), (2) Platform as a Service (PaaS), and (3) Infrastructure as a Service (IaaS).
Among the three models, IaaS is the most abused model by the attackers. The SaaS (e.g., DropBox) model offers users with minimal customization options; thus it is difficult to abuse. PaaS (e.g., Google App Engine\footnote{https://cloud.google.com/appengine/}) enables users to deploy their applications on cloud, however, using API restrictions, misuse of PaaS model can be prevented. IaaS model empowers the users with extreme flexibility. The enormous processing power and storage capability provided by the IaaS cloud at a minimum cost enable cyber attackers to conduct a plethora of malicious activities using the cloud. The cyber attackers also take advantage of the weak authentication and monitoring capabilities on the cloud that does not require them to put much effort into hiding their tracks.\\ 
\textbf{Hosting Phishing Websites on the Cloud:} 
Attackers are now able to host a phishing website on the cloud platform to steal credentials of legitimate users on the Internet \cite{Modi2013B, MISHRA201718, c15, c15B}. 
Attackers had developed a phishing website that asked users to enter their Microsoft 365 credentials\footnote{https://www.cyren.com/blog/articles/point-click-andhack-phishers-try-wix}\textsuperscript{,}\footnote{https://www.infoworld.com/article/3187346/phishingscammers-exploit-wix-web-hosting.html}. The website was designed and hosted on a popular website creation and hosting service called {\tt http://www.wix.com}. The wix website was designed to mimic a login page on Microsoft's website to trick the unsuspecting users into giving away their credentials. Hosting such phishing websites is a cost-effective way for attackers instead of paying for the expensive physical resources that might traced back to them.
\\
\textbf{Cloud as a Media to Launch DoS/DDoS Attacks:}
As a general strategy, attackers are always trying to find novel ways to launch cyber attacks. One such example is the attacker hosting a botnet on a cloud to launch a DDoS attack, as in the case of Zeus botnet being hosted on Amazon's EC2 cloud services \cite{choo2010cloud}. In addition to using botnets, attackers can also use various freely available tools such as Low Orbit Ion Cannon (LOIC) installed on the cloud to launch DoS attacks \cite{Mike2012}. 
In addition, LOIC offers a web-based tool to launch the attacks from within the browser without needing to install anything. LOIC can launch packet-flooding attacks using HTTP, TCP, and UDP packets. It has now become a popular choice for attackers for DoS/DDoS attacks after becoming open-source\footnote{https://github.com/NewEraCracker/LOIC}.
These tools can stealthily scan for open ports and services on an IP address and then use them to flood the ports with messages and  launch a DoS attack.\\
\textbf{Brute Force Attacks:} 
In 2010, Amazon officially announced that its AWS website received some user reports of SIP (Session Initiation Protocol) brute force attacks originating from Amazon EC2. SIP brute force attack most commonly uses vulnerabilities in SIP protocol for password auditing in VoIP (Voice over IP) sessions through brute force attack\footnote{https://aws.amazon.com/security/security-bulletins/sip-abuse/}. According to a study \cite{choo2010cloud}, if an attacker wants to use Amazon EC2 to brute force a 10-character password, which contains only lower-case letters, it would cost the attacker less than US \$2,300 based on the price Amazon asks for an hour of EC2 web service usage.\\
\textbf{Rogue Cloud:} 
Cyber attackers might take advantage of cloud computing to offer services, especially in regions that suffer from a lack of cyber crime laws and regulations. These rogue cloud services which provide hosting and data services for a lower price can be used for criminal purposes such as objectionable or copyrighted contents and, at the same time, can be hidden from law enforcement authorities. Charging a lower hourly fee, these rogue cloud services are also options (i.e., honeypots) for less aware clients who risk the leakage of their data \cite{choo2010cloud}.\\
\textbf{Generic Attacks:} 
Many malicious activities can be performed by abusing the cloud services including 1) password and key cracking, 2) intrusion attacks, 3) port scanning, 4) sending spams, 5) launching dynamic attack points, 6) hosting or distributing malicious software, 7) botnet command and control, 8) building rainbow tables which stores the hashes of large number of strings, and 9) \textit{CAPTCHA} solving farms, which solve the captchas in exchange of pay. It should be considered that cloud service providers always declare that these attacks are not specific only to the cloud services but could also be launched from any computer connected to any network \cite{choo2010cloud}.

\section{Case Study}
\label{sec:experiment}
This section provides the details of the replicating the steps taken by the attackers on a simulated and controlled environment and reports the details of the detection mechanisms and results. Our main focus is to highlight the groundwork for \textit{``Proactive Forensics''} of the attack VMs, so that they can be identified and isolated before an exploitation.

\subsection{Platform Setup and Simulation Details}
We setup two different VMs on the Oracle VirtualBox. One of the VMs was designated as the attack VM, while the other was considered to be the target VM. The \textit{attack VM} was a Debian-derived Linux distribution Kali Linux\footnote{https://www.kali.org/} as it comes with the necessary tool set for performing steps to launch an attack. The target VM ran Windows 7 and was used for port scanning and propagating a malware.  

The primary reason behind having VMs as simulated environment is to replicate the abstraction of the physical devices provided in the virtualized environment of cloud. The only difference between the cloud virtualization and the virtualization utilized for the simulation of this work is the type of hypervisors that are used. Cloud infrastructures use the \textit{Type 1 Hypervisor} (i.e., \textit{Virtual Machine Manager}), that runs directly on the hardware platform; Whereas, we have simulated the case study using \textit{Type 2 Hypervisor}, that runs on a host OS. 

We collected various log files (e.g., guest OS logs, host and guest application logs, firewall logs) of both the attack VM and the target VM, as we performed the malicious activities, 
to identify the traces of those activities. The aim of this case study is to show that it is possible to perform live forensics to identify when a VM is used for launching an attack. Hence, 
the various log files can be useful indicators for maliciousness. 

\subsection{Attack Scenarios}

\subsubsection{Suspicious Activity Scenario \#1 (Port Scanning)}
All the interview responses that the research team collected through the survey questions almost invariably reported port scanning as a popular choice of reconnaissance. The cyber attackers employ port scanning for numerous reasons and the responses we received indicated that the port scanning is performed to identify the vulnerabilities or blocking functionalities of the target system or it is used as a way to leave a backdoor for launching further attacks. We used the Nmap tool\footnote{https://nmap.org/} to perform the port scans. 

Our attack VM (Kali Linux) and the target VM (Windows 7) were running on the same physical machine. The first step was to find out the IP addresses of both VMs. The commands \textit{\textbf{ifconfig -a}} and  \textit{\textbf{ifconfig -eth0}} can capture the IP addresses of the VMs. Once we have the attack VM IP, using the Nmap's  IP range scan command \textit{\textbf{nmap -sn 10.0.2.1-255}}, we obtained the other live VMs that can be potential targets. 

 After obtaining the IP addresses of the target VM, we performed port scans using Nmap. Nmap provides various functionalities to perform a scan. We scanned a range of IPs on the target VMs. 

\subsubsection{Suspicious Activity Scenario \#2 ( Malicious Executable)}

The goal for this activity is to capture the traces of malicious code propagation using VM. We implemented the scenario using a Windows malware\footnote{MD5: e5dce3d5e39a5e790a407c3e0632b887} and set up a clean Windows 7 VM for running the malicious executable. We disabled the Windows Defender Services, Windows Security Services, Firewall and other automatic security updates, so that a malware can run uninterrupted on the VM, 
We let the malware run for 2 minutes on the VM and captured the execution event using Process Monitor tool\footnote{https://docs.microsoft.com/en-us/sysinternals/downloads/procmon}.

The malware sample, a ransomware, in the PE executable format was obtained from VirusShare\footnote{https://virusshare.com/}. The malware interactions with (1) file system, (2) registry system, (3) API calls, (4) network  and (5) processes were captured using Process Monitor tool. By organizing appropriate filters on Process Monitor, the tool can capture the run time behavior of the malware. We then saved the output in a CSV (Comma Separated Value) file and used for further analysis. 

\subsubsection{Suspicious Activity Scenario \#3 (Denial of Service)}
\begin{figure}[h]
  \includegraphics[width=\linewidth]{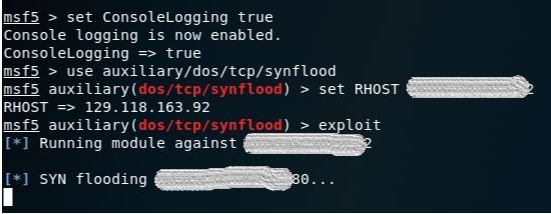}
  \caption{TCP SYN flood using Metasploit.} 
  \label{fig:meta}
  \vspace{-0.15in}
\end{figure}

Denial of Service (DoS) attack can be simulated at different levels: (1) Application based, that targets to exhaust the target OS resources, (2) Protocol based, that exhausts the connection pool of the target, and (3) Volume based, that floods the network bandwidth of the  target. We simulated a Protocol based DoS attack on one of our internal server (Dell PowerEdge T630), using the attack VM. We utilized the open-source penetration tool framework Metasploit\footnote{Integrated into Kali Linux: https://www.metasploit.com/}. It is built-in into the Kali Linux of our attack VM. Figure \ref{fig:meta} shows the screenshot of using metasploit for launching DoS attacks. 

As shown in Fig. \ref{fig:meta}, we launched the Metasploit framework by executing the \textbf{\textit{msfconsole}} command. We selected the auxiliary \enquote{\it auxiliary/dos/tcp/synflood} for performing the TCP SYN flood on the target server. Once the auxiliary was loaded, we set the \textbf{\textit{RHOST}} to the IP address of our target internal server.

\subsection{Results}
\label{sec:results}

\subsubsection{Port Scanning}
The port scan activities performed by Nmap were not easily identifiable from the system logs obtained from the target VM. The Intrusion Detection Systems (IDS) are a popular choice to spot the port scans activities on target machines. However, attackers can customize the scanning rules through the Nmap Data Files and perform the scan discreetly to stay undetected. The firewall logs, IDS logs, and system logs can show the trace of a port scan on a target system but these logs are generally huge and are often not subjected to thorough inspection. 

\begin{figure}[h]
  \includegraphics[height=4cm,width=\linewidth]{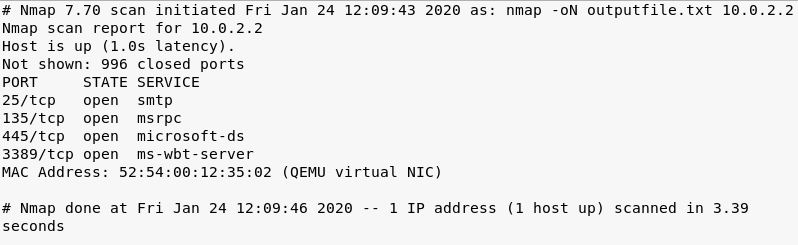}
  \caption{Snapshot of Nmap log during a port scan.} 
  \label{fig:nmap}
  \vspace{-0.15in}
\end{figure}

To enumerate the information that can be captured while a port scanning takes place on the attack VM, we captured the Nmap log. Figure \ref{fig:nmap} is the log snapshot of a port scanning activity performed using the Nmap.

The log snapshot shown in Figure \ref{fig:nmap} indicates:
\begin{itemize}
    \item The timestamp of initiation of the scanning task
    \item List of all the open ports
    \item The protocols running at each port
    \item The services running on each of the open ports
    \item The physical or MAC address of the target machine
    \item The total time taken up by the Namp to finish the scan
\end{itemize}

If these informative data are captured and analyzed on the attack VM, it is possible to identify when a virtual machine (or in a cloud perspective a computing instance) is abused for performing a port scan.

\subsubsection{Malicious Executable}
\begin{figure}[h]
  \includegraphics[height=3cm, width=\linewidth]{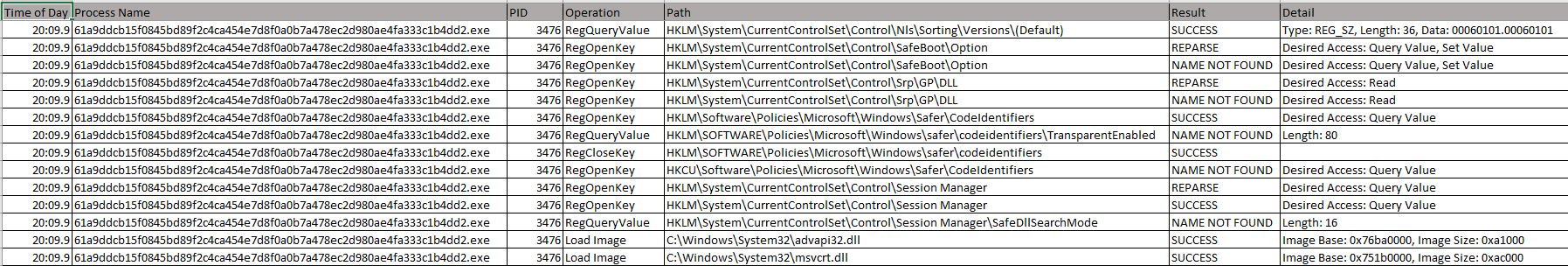}
  \caption{Log file entries from running a malicious executable.}
  \label{fig:mal}
    \vspace{-0.15in}
\end{figure}

Figure \ref{fig:mal} presents a snapshot of dynamic behavior of the ransomware sample mentioned in Section \ref{sec:experiment}. As the ransomware executes, we captured its dynamic behavior using the Process Monitor tool. The results show the interaction of the malicious executable with the system registry and other processes. 

The registry operations are essential in understanding the persistence mechanism of the malware; whereas, the network activities look for the connection attempts made by the malicious executable. The file operation identifies the created, deleted and modified list of files by the malware. It also shows the API calls sequences, process interaction attributes to the identification of the purpose of the malware 
The log provides some indicators of malicious activities.
According to the VirusTotal\footnote{https://www.virustotal.com} scan results, the malware is an encrypting ransomware. From the log in Figure \ref{fig:mal} we observe that the ransomware carries out registry operations to locate system files. A few other entries from the Process Monitor log show that the malware: (1) Opens and parses various DLL (Dynamically Linked Libraries) files. (2) Accesses the Internet Explorer (IE) cookies and other saved information. (3) Accesses various application and setting data that can help the running malware to identify virtual environment. With the information captured by the Process Monitor tool, automated observation methodologies can be utilized to tag the malicious activities taking place on the virtual or cloud environments.

\subsubsection{Denial of Service}

\begin{figure}[h]
  \includegraphics[height=5cm,width=\linewidth]{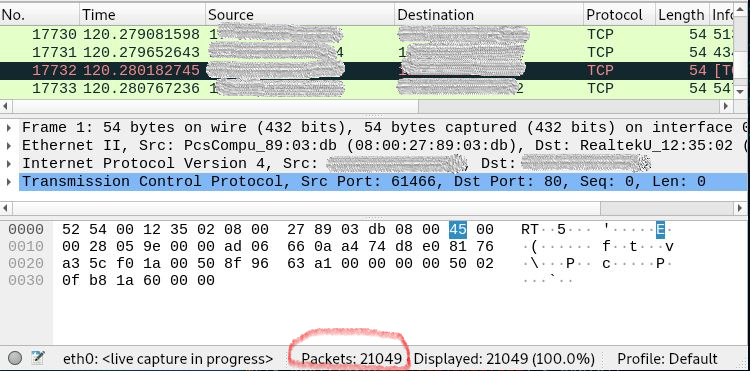}
  \caption{Wireshark statistics.} 
  \label{fig:wire}
  \vspace{-0.15in}
\end{figure}

We used Wireshark\footnote{https://www.wireshark.org/} to capture the packet flow in order to identify whether a DoS attack occurred. Instead of placing the Wireshark on the target VM, we ran it in the attack VM and captured the number of packets it sent out. While TCP SYN flood was the only network activity happening on the attack VM, the Wireshark interface showed a packet volume of $21,049$ within a few seconds of launching the exploit, as highlighted in red in Figure \ref{fig:wire}. 

\subsection{Sources of Evidence}
In this section, we enumerate the information that are available through various log files that can be used as evidence to identify potential abuse on a virtual environment and discuss the limitation and challenges involved in virtual disk forensics. To execute forensic activities, we need to prepare, acquire, preserve, analyze, and report the anomalies in a timely manner. 
For traditional forensics, we have physical data to analyze. However, in virtual environments it can be inconvenient and also hard to acquire evidence. This section lists the forensic attributes available for Oracle VirtualBox, which enables virtualization on a local computer.
For the purpose of VMs, mounting the snapshot of the attack VM to a host and analyzing the relevant files/processes inside it provides useful insights. 
\subsubsection{Virtual Machine Snapshots}
The forensics on the VM images requires contextual data (i.e., dynamic configuration). Therefore, different files generated by the hypervisor needs to be analyzed and monitored. For the case study conducted in this paper, the following files were generated for the Kali Linux, Ubuntu and Windows 7 VMs on Oracle VirtualBox: \\
\textbf{1) sav Files.} These files are the memory-content when the virtual machine state is saved. However, they cannot be utilized to replicate the hard drive. These snapshot files are used for replications.\\
\textbf{2) vdi Files.} These types of files are the container format for guest hard disks. This format is recognized by Oracle VirtualBox, and it is created anytime a new host VM is created.\\
\textbf{3) log Files.} These log files contain detailed configuration and runtime information about each virtual machine.\\
\textbf{4) XML files.} These files describe the settings of the VM in an XML format.

\subsubsection{Virtual Hard Drive}    
The virtual hard drives are the files containing the complete content and structure pertaining to the underlying virtual hard disk drive.  They are used to store the virtual OS and the related information. In fact, these files are the results of simulations of a physical hard drive. A VHD of the VM can be mounted and analyzed for forensics purposes.

\subsubsection{Virtual Machine Forensics}
The binaries of the attack VHDs can be difficult to achieve an optimized analysis. On the other hand, the capturing and restoring the attack VM from its snapshot is easier to find the evidence of anomalous behavior.

\subsubsection{Limitation and Challenge}
Virtual log files along with the host file systems could have been encrypted and therefore, for forensics analysis the files need to be decrypted. The cloud platforms provide users with large amount of resources, memory, and processing resources. Capturing and analyzing that large amount of resources for forensics needs time and further processing power. Any user on the cloud can be a potential adversary. It is practically infeasible to predict which user could potentially abuse the cloud-based resources. To prevent the abuse of cloud platform, network traffic and resource utilization for each user needs to be monitored. More precisely, it is suggested that both adversaries and benign users are put under surveillance, compromising the privacy of the individuals. Hence, defining the amount of perusal for individual is crucial and subject to defining security policies. 

\section{Related Work}
\label{sec:relwrk}

Addressing the incident response and forensics investigation on IaaS cloud platform, Dykstra et al. \cite{frost} proposed a forensics platform to capture the virtual disks from OpenStack\footnote{https://vexxhost.com/private-cloud/} cloud platform. The proposed platform FROST attempts to operate at the management plane of the cloud. Through API level function calls, FROST enables the user to capture evidences from virtual disks through API and firewall logs. 

The challenges and short comings for forensics  activity in heterogeneous paradigm of Internet of Things, Zawoad et al. \cite{faiot} presented \textit{Forensics-aware IoT} (FAIoT) model. The FAIoT model combines cloud forensics, network forensics and device forensics to address the complexity of identification, collection, organization and presentation of the relevant IoT forensic evidences. 
In another work, Zawoad et al. \cite{laas}  Secure-Logging-as-a-Service (SecLaaS) for enabling cloud forensics. Since every  user activity on cloud can be traced from activity and other logs,  SecLaaS scheme securely store the log files in a persistence database and creates an entry in the \textit{proof-database} for each log. These database entries ensure that the log files are available even when the VM is terminated on the cloud, the logs for a particular IP is available to the investigators. 

\section{Conclusion and Future Work}
\label{sec:conclude}

With the advent of the Internet and increasing cost of computing resources \cite{c1}, cloud computing has become the most appealing computing paradigm that provides  resources as a utility. These features are specifically attractive for attackers as they can use the functionality of cloud and can still remain covert. In this research work, we interviewed over 75 professional and ethical hackers with the aim of understanding their mindset when conducting cyber attacks. The participants were allowed to explore their hacking experience with respect to a set of hypothetical attack scenarios. While analyzing their responses and building a mental model to structure their mindset, it was observed that professional hackers heavily utilize cloud for different purposes including masking their identities and utilizing computational powers to launch DDoS attacks or create a botnet. Inspired by the output of the interviews, we identified the types of cyber attacks often launched using cloud. We performed a number of case studies to simulate possible attacks that can be launched through cloud environment. Our case studies targeted 1) port scanning, 2) malicious execution, and 3) denial of service attacks. We then captured some log files and analyzed them in order to demonstrate the feasibility of tracing attacks through log file analysis of such cloud-based platforms. 
Our research needs further work on building a forensics suite and also developing security testing framework \cite{c20} for the VHD images that would enable real time forensics on the VM instances on cloud platforms. The cloud abuse can also be prevented by employing formal adaptive security techniques and in the presence of uncertainty \cite{SaraSAC2017}. One of the challenging problems is how to detect cloud abuse without any historical data (i.e., ``{\it zero-day cloud abuse}''). The problem is very similar to detection of zero-day malware \cite{c30}. However, it requires developing specific techniques in cloud.

  \vspace{-0.01in}
\section*{Acknowledgment}
Thanks Sara Sartoli for her contribution to the interview data collection. This research work is supported by National Science Foundation under Grants No: 1516636, 1723765, 1821560.

\vspace{12pt}


\vspace{-0.01in}
\begin{thebibliography}{00}
\bibitem{c9} Cloud Security Alliance, \enquote{The Treacherous 12 - Top Threats to Cloud Computing + Industry Insights}, (2017), \url{https://downloads.cloudsecurityalliance.org/assets/research/top-threats/treacherous-12-top-threats.pdf}
\bibitem{c2} Microsoft Security Intelligence Report, Volume 22, January through March, 2017.
\bibitem{c30} Faranak Abri, Sima Siami-Namini, Mahdi A. Khanghah, Fahimeh M. Soltani, and Akbar S. Namin, Can machine/deep learning classifiers detect zero-day malware with high accuracy? In IEEE Big Data, 2020.
\bibitem{c11} P. Arntz. How to protect your RDP access from ransomware attacks, (2018). [Online]. Available: \url{https://blog.malwarebytes.com/security-world/business-security-world/2018/08/protect-rdp-access-ransomware-attacks/}
\bibitem{c15-0} Moitrayee Chatterjee, Prerit Datta, Faranak Abri, Akbar Siami Namin, and Keith S. Jones, Abuse of the Cloud as an Attack Platform, IEEE COMPSAC, 2020. 
\bibitem{c15} Moitrayee Chatterjee and Akbar Siami Namin, Detecting web spams using evidence theory
 IEEE COMPSAC, 2018. 
 \bibitem{c15B} Moitrayee Chatterjee and Akbar Siami Namin, Detecting Phishing Websites through Deep Reinforcement Learning, COMPSAC, 2019.
\bibitem{choo2010cloud} K.K.R. Choo, Cloud computing: challenges and future directions. Trends and Issues in Crime and Criminal justice, (400), p.1., 2010
\bibitem{Mike2012} M. Danseglio, Ethical hacking: How to create a
dos attack, 2012. [Online]. Available: \url{https://www.pluralsight.com/blog/itops/ethical-hacking-how-to-create-a-dos-attack}
\bibitem{c20} Shuvalaxmi Dass and Akbar Siami Namin, Vulnerability Coverage for Adequacy Security Testing, In ACM SAC, 2020.
\bibitem{frost} J. Dykstra and A.T. Sherman, 2013. Design and implementation of FROST: Digital forensic tools for the OpenStack cloud computing platform. Digital Investigation, 10, pp.S87-S95.
\bibitem{cloudnist} P. Mell and T. Grance, “The NIST Definition of Cloud Computing,” NIST, Special Publication 800-145, September 2011.
\bibitem{MISHRA201718} P. Mishra, E.S., Pilli, V. Varadharajan and U. Tupakula, Intrusion detection techniques in cloud environment: A survey. Journal of Network and Computer Applications, 77, pp.18-47. 2017.
\bibitem{Modi2013B} C. Modi, D. Patel, B. Borisaniya, A. Patel and M. Rajarajan, A survey on security issues and solutions at different layers of Cloud computing. The journal of supercomputing, 63(2), pp.561-592, 2013.
\bibitem{SaraSAC2017} Sara Sartoli, Akbar siami Namin,
A semantic model for action-based adaptive security, ACM SAC, 2017.
\bibitem{laas} S. Zawoad, A.K. Dutta, and R. Hasan, 2015. Towards building forensics enabled cloud through secure logging-as-a-service. IEEE Transactions on Dependable and Secure Computing, 13(2), pp.148-162.
\bibitem{faiot} S. Zawoad and R. Hasan, 2015, Faiot: Towards building a forensics aware eco system for the internet of things. In 2015 IEEE International Conference on Services Computing (pp. 279-284). IEEE.
\bibitem{c1} S. Zhang, S. Zhang, X. Chen, X. Huo, \enquote{Cloud Computing Research and Development Trends}, Int. Conference on Future Network, 2020.
\end{thebibliography}
\end{document}